\documentclass[12pt,epsf]{article}
  \usepackage{epsfig}
  \newlength{\abstractwidth}
  \renewcommand{\thefootnote}{\fnsymbol{footnote}}
  \renewcommand{\thanks}[1]{\footnote{#1}} 
  \newcommand{\starttext}{
  \setcounter{footnote}{0}
  \renewcommand{\thefootnote}{\arabic{footnote}}}
  \renewcommand{\theequation}{\thesection.\arabic{equation}}
  \newcommand{\be}{\begin{equation}}
  \newcommand{\bea}{\begin{eqnarray}}
  \newcommand{\eea}{\end{eqnarray}}
  \newcommand{\beq}{\begin{equation}}
  \newcommand{\ee}{\end{equation}}
  \newcommand{\eeq}{\end{equation}}

  \def\ba{\begin{eqnarray}}
  \def\ea{\end{eqnarray}}

\def\14{{1\over4}}
  \def\12{{1 \over 2}}
  \def\eq{&=&}
  \def\tro{\tilde{\rho}}
  \def\dro{\delta{\rho}}

  \def\h3{h^{3\over 2}}

  \def\lb{\label}
  \def\f{\phi}

  \def\V{V(\phi)}
  \def\epg{{8 \pi G \over 3}}

  \def\CC{cosmological constant }
  \def\delt{\delta \rho/\rho}
  \def\simleq{\; \raise0.3ex\hbox{$<$\kern-0.75em
      \raise-1.1ex\hbox{$\sim$}}\; }
   \def\simgeq{\; \raise0.3ex\hbox{$>$\kern-0.75em
      \raise-1.1ex\hbox{$\sim$}}\; }

\begin{document}
  \renewcommand{\theequation}{\thesection.\arabic{equation}}
  \begin{titlepage}
  \bigskip

  \bigskip\bigskip\bigskip\bigskip

  \centerline{\Large \bf {Observational Consequences of a Landscape}}

\vspace{20pt}
\begin{center}
Ben Freivogel${}^1$, Matthew Kleban${}^{2,4}$, 
Mar\'{\i}a Rodr\'{\i}guez Mart\'{\i}nez${}^{3,4}$ and
Leonard Susskind${}^1$

\vspace{25pt}

${}^1$ {\it Department of Physics,
  Stanford University, Stanford, CA 94305-4060, USA}
  
\vspace{10pt}

${}^2$ {\it Institute for Advanced Study, Princeton, NJ 08540, USA}

\vspace{10pt}

${}^3${\it The Hebrew University, Jerusalem 91904, Israel}

\vspace{10pt}

${}^{4}${\it Department of Physics and Astronomy, University of Texas, Austin, TX USA}

\end{center}
\vspace{10pt}

  \bigskip\bigskip
  \begin{abstract}

In this paper we consider the implications of the ``landscape"
paradigm \cite{Bousso:2000xa, landscape} for the large scale properties of the universe. The most
direct implication of a rich landscape is that our local universe
was born in a tunnelling event from a neighboring vacuum. This
would imply that we live in an open FRW universe with negative
spatial curvature. 
We argue that the ``overshoot" problem, which in other settings would
make it difficult to achieve slow roll inflation, actually favors such a cosmology.

We consider anthropic bounds on the value of the curvature and
on the parameters of inflation. When supplemented by statistical
arguments these bounds suggest that the number of inflationary
efolds is not very much larger than the observed lower bound.
Although not statistically favored, the likelihood that the number
of efolds is close to the bound set by observations is not
negligible. The possible signatures of such a low
number of efolds are briefly described.

  \medskip
  \noindent
  \end{abstract}

  \end{titlepage}
  \starttext \baselineskip=17.63pt \setcounter{footnote}{0}


\setcounter{equation}{0}
\section{Introduction}

In what follows we will assume the validity of what has come to be
called the ``Landscape Paradigm" for cosmology. The importance of
observationally testing this hypothesis is obvious. In this paper
we will discuss the consequences of one particular
aspect of the paradigm--the requirement that our ``pocket
universe" was born in a tunnelling event from some neighboring
valley of the landscape.

By the Landscape Paradigm we will mean three things \cite{Bousso:2000xa, landscape}:

\begin{itemize}
\item The string theory landscape of metastable de Sitter vacua is extremely rich. So many vacua exist
that the large numbers can compensate the apparent fine-tuning of the cosmological constant, the gauge
hierarchy, and whatever additional fine-tunings are phenomenologically required by observational data. Specifically we
assume a large set $\cal {S}$ of minima consistent with the
standard model and the small measured cosmological constant.

\item The global universe is an eternally inflating ``megaverse" that is continually
producing pocket universes (from now on, just universes) by tunnelling events. Our own universe
went through a series of tunnelling events, the last of which led to a conventionally inflating universe
that eventually settled into a valley of the landscape with the tiny vacuum energy we see today. The rest of the
megaverse is filled with universes that populate the entire landscape.

\item Some features of the observed laws of physics are contingent on our environment--our location in the landscape.
The answers to some questions will be anthropic: life exists only where the conditions are right.
For example, the cosmological constant is small in our universe because if it were not, we wouldn't be here to ask about it.

\end{itemize}

We emphasize that these three items are not arbitrary speculations. String theory  is the only framework
that we know of in which quantum mechanics and gravity coexist in a consistent mathematical framework.
The string theory landscape is the mathematical consequence of applying standard low energy approximations to string
theory. As for the second item, eternal inflation and bubble nucleation appear to be the consequences of an initial large energy
density on a landscape of de Sitter vacua. A generic starting point will lead to the universe,
 or some patch of it, getting stuck in one of the many local minima of the potential.
 Once that happens, eternal inflation takes over.
 Finally, given the first two items, the last is no more than a tautology: we live where life is possible.

The most important cosmological consequence of these assumptions is that there was a tunnelling event in our past.
A universe that is born by tunnelling, from a neighboring valley of the landscape, has one distinguishing characteristic
that is the focus of this paper: that universe is an \it infinite open FRW universe \rm with negative spatial curvature and with very
special initial conditions \cite{Coleman:1980aw}. The best short-term hope for confirmation of the Landscape Paradigm may come from the special features of such a cosmology.

The  paper consists of four parts. In the next section (section 2) we will review the characteristics of a universe born
out of a tunnelling event. We will sketch out the early history of such a universe from the tunnelling event until
the universe enters a conventional slow-roll inflationary phase.

One very important question  is why the universe tunnelled into a conventionally  inflating vacuum with a large
number of efolds. This evolution is not generic and requires a good deal of fine-tuning. In section 3 we derive
anthropic constraints on the curvature of the universe at decoupling. These constraints are translated into constraints
on the amount of inflation in section 4. We also use statistical
arguments of the kind pioneered by Douglas and collaborators \cite{douglas} \cite{douglas2},
to estimate the duration of inflation, i.e., the
number of efolds. Our statistical analysis is suggestive but preliminary.
A full analysis would require knowing the string theory landscape and also
understanding the proper measure on the landscape.  We find that the number of efolds is unlikely to be very much larger than the observed lower bound.
In fact the probability that the
number of efolds is within two or three of the observed bound is surprisingly large, of order ten percent.

Finally, in section 5 we discuss the empirical consequences of a small number of efolds. These include measurable
spatial curvature and deviations from scale invariance for the lowest modes of the CMB fluctuation spectrum.

{\bf Summary of Results.} Our results are most simply stated in terms of the number of efolds. Assuming standard inflationary parameters, we find that 59.5 efolds are required for structure formation. The current observational bound on the curvature requires at least 62 efolds\footnote{These numbers are weakly dependent on such parameters as the reheating temperature.  However, varying them would not significantly alter the results stated here.}. We use a crude parameterization of the inflationary potential to find the probability distribution for the number of efolds $N$, finding that it is proportional to $1/N^4$. The anthropic bound together with our probability distribution give a probability of about 90\% that the number of efolds is greater than 62, so the statistical argument is consistent with observation. Given the observed bound of 62 efolds, we find a probability of about 10\% that the actual number of efolds is between 62 and 64. Such a small number of efolds would have observable consequences.

{\bf Relation to Other Work.}
Some of the results we derive here are not new. Open inflation from bubble nucleation was discussed in \cite{gott, ratra} among others.  Barnard and Albrecht \cite{albrecht} discussed the landscape and open inflation. Tegmark \cite{teg} modelled the landscape and its predictions. The CMB data, along with consequences of spatial curvature, have been discussed for example by \cite{lowlanomaly}. Open inflation, its observational consequences, and the probability distribution for the number of efolds was discussed in various frameworks by \cite{HTetc}. The anthropic bound on the curvature for an open universe seems to have been derived first by Vilenkin and Winitzki \cite{vilenkin} and discussed in a context similar to ours by Garriga, Tanaka, and Vilenkin \cite{garr}.
The measure problem in eternal inflation, which needs to be solved in order to do a correct statistical analysis, has been discussed in \cite{measure}. We consider the measure problem to be an important open problem, and we simply ignore the possible volume factors in the measure. Possible measures and their paradoxes are discussed in \cite{measure}.

\setcounter{equation}{0}
\section{The Universe After Tunnelling}

For the purposes of this paper we are only interested in the part of the landscape in our immediate
vicinity. We will model our neighborhood by a single scalar inflaton field and a potential function
that includes
three relevant regions. The
first is a metastable minimum with a relatively large vacuum
energy and a large curvature (mass). This region is separated from
the region of conventional inflation by a high barrier. Tunnelling from this
minimum leads to
 a broad, flat plateau with a very shallow slope that
eventually gives way to a steep reheating slope. The steep slope
leads to today's minimum with a tiny cosmological constant. The
portion of cosmological history that will concern us begins with the
tunnelling event from the initial minimum to the inflationary
plateau, continues with slow roll inflation to the reheating slope, and finishes with conventional Big
Bang cosmology up to the present era.

\begin{figure}
\begin{center}
\includegraphics{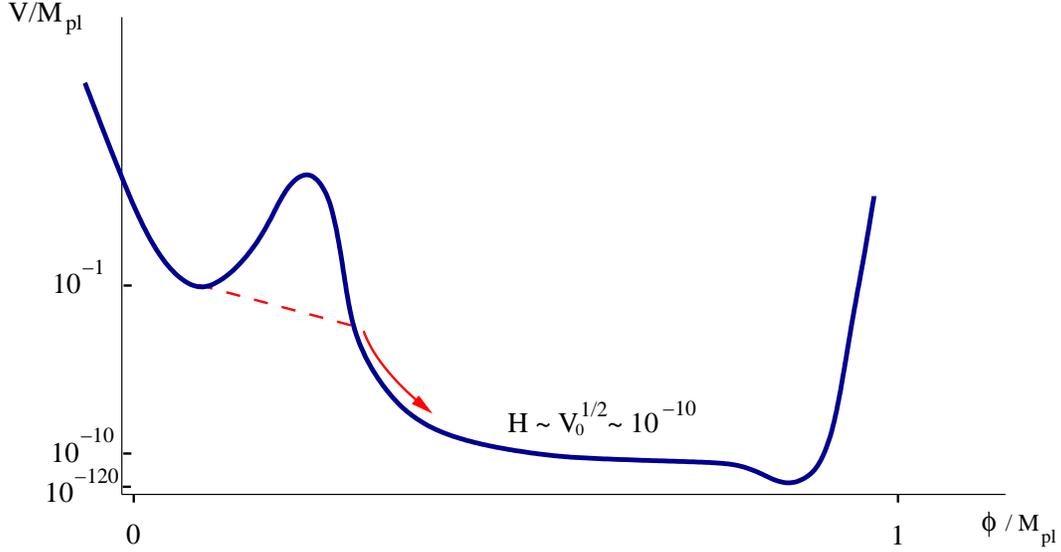}
\caption{A highly unusual potential which is a typical example of the ones we will consider.}
\label{pot}
\end{center}
\end{figure}

The vacuum in the left-most minimum is obviously metastable and will decay by bubble nucleation.
The theory of this process, developed long ago by Coleman and De Luccia \cite{Coleman:1980aw}, is
described by an instanton--a solution of the Euclidean equations of motion--with the topology
of a four-sphere. The instanton preserves an SO(4) out of the SO(5) symmetry of the four-sphere.

The continuation to Lorentzian signature describes an expanding bubble in the metastable
de Sitter background. The metric in the bubble is given by an FRW open geometry with metric
\bea
ds^2 \eq dt^2 - a(t)^2 dH_3^2 \, , \cr
dH_3^2 \eq dr^2 + \sinh^2{r} \ d\Omega^2 \, ,
\lb{frw}
\eea
where $a$ is the scale factor, $dH_3^2$ is the metric of a uniformly negatively curved space with unit
curvature, and $d\Omega^2$ is the metric of a unit two-sphere. The spatial hypersurfaces are
homogeneous negatively curved spaces with curvature $1/a^2$. The original
SO(4) symmetry of the instanton becomes the noncompact SO(3,1) of the hyperbolic spatial slices.

\begin{figure}
\begin{center}
\includegraphics[width=16cm]{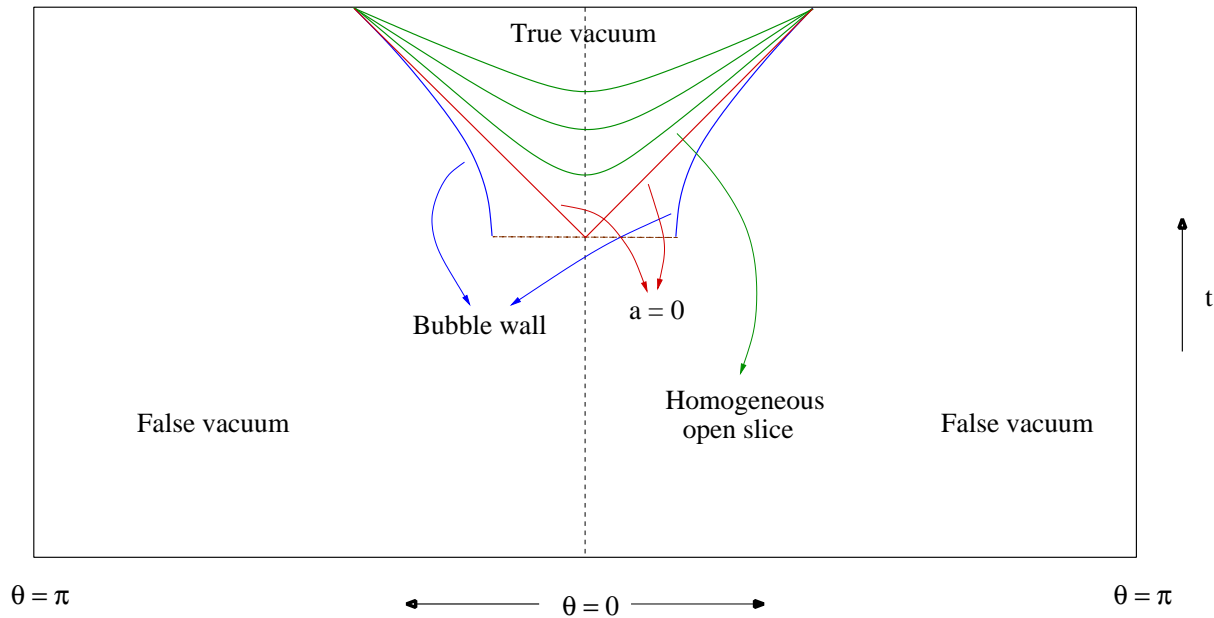}
\caption{A conformal diagram showing the bubble nucleation.  The background space, labelled ``False vacuum," is the initial de Sitter minimum. }
\label{penrose}
\end{center}
\end{figure}

The equations of motion are
\bea
({\dot a \over a})^2 \eq {8 \pi G \over 3}\left( {\dot{\f}^2 \over 2 } +\V    \right)+{1 \over a^2}     \cr
\ddot{\f} + 3H\dot{\f} \eq -V'(\f)
\lb{eqmot},
\eea
where $H = {\dot{a} / a}$.

The initial conditions at $t=0$ for this FRW $k=-1$ universe are
very special, non-singular initial conditions that reflect the
smoothness of the instanton. They are: 
\bea a(t) \eq t + {\cal O}(t^3) \cr 
\dot{\f}(0) \eq 0 
\lb{bcs} 
\eea 
We choose $\f(0) = 0$ for convenience.  Close to $t=0$ the scale factor behaves like $a(t) \sim
t$. To see that this is smooth, note that flat space can be
written in the form (\ref{frw}) with $a(t) = t$, so $a=0$ is simply
a coordinate singularity. The second equation gives the
initial condition for $\f$. The field starts at rest at the point
where it emerges from the tunnelling event. This follows from the
symmetry of the instanton and is also required by the equations of
motion.

For a period of time the equation of motion for $a$ is dominated by the curvature term $1/a^2$ and one finds
\bea
a(t) \eq t \cr
{\dot{a} \over a} \eq {1 \over t} = H.
\lb{cv}
\eea
The friction term in (\ref{eqmot})  diverges as $t \to 0$, and consequently, $\f $ moves slowly at first.

Eventually the curvature term will fall below the potential energy and the equations will become dominated by the
effective cosmological constant on the plateau, $V(plateau)= V_0$. This happens at a time that we call $t^*$,
\be
t^*= \sqrt{{3 \over 8\pi G V_0}}
\lb{tstar}.
\ee
Once $t$ passes $t^*$ the universe begins to inflate while it slowly rolls down the plateau.

\subsection{An overshoot problem?}
A potential problem with inflationary cosmology was pointed out by Brustein and Steinhardt \cite{ram} (see also \cite{Kaloper:1991mq}). The value of the
inflaton potential during inflation must be very small in Planck or string units: most likely smaller than $10^{-14}$.
This means that if the inflaton descends from some string scale energy density it will generally overshoot the inflationary
region unless the potential descends extremely gradually. Thus it is usually necessary to make the range of
the inflaton field to be many Planck units. Such a large range is questionable, and difficult to arrange in string theory.
As we will see below, the overshoot problem simply does not occur in a theory with a tunnelling origin.

To understand the overshoot problem and its solution, assume that
 the field starts at a high value of the potential relative to the inflationary plateau.
It will have a large velocity by the time it reaches the plateau (see Figure \ref{pot}).
If the only friction in the equation of motion on the plateau were due to the vacuum
energy $V_0$, the field would shoot across the plateau and the universe would reheat with no period of inflation.

In fact, this is not at all what happens in a universe that evolves from a
tunnelling event.  The  curvature provides a large friction term which easily checks
the motion of the inflaton.
Generically the distance the field travels along the plateau before slowing to slow-roll speeds
is independent of the initial height where it begins!  We can illustrate this with a simple model.

Suppose the potential consists of two linear sections. The initial
section is very steep, and covers the region in field space $0 <
\phi < f$ where $f$ is assumed to be a small fraction of the Planck mass.
We will take $V(0) = V$, so that the slope in the
initial section is $-V/f$. The field begins at rest at $\phi=0$.
The next section ($\phi > f$) represents the plateau, which for
this analysis we take to be flat and with an energy small enough that for a long period it can be ignored. We assume curvature
domination throughout\footnote{An  alert reader may be
concerned with the approximation of curvature domination.  In
fact, as long as $f<1$ the kinetic field energy never dominates the curvature term.}, so the equation of motion is
\be
\ddot \phi
+ 3 \dot \phi/t - V/f=0
\ee
in the first section, and the same without the $V/f$ term in the
second. Solving the equations and matching shows that
\be
\phi(\infty)=2f,
\ee
so the distance travelled along the plateau is
simply $f$, independent of the initial height $V$. This surprising
result indicates that no particular fine-tuning is necessary to
avoid overshoot. The potential energy only dominates
long after the field velocity has become negligible.
\subsection{Horizon evolution}

Let's consider the history of the Hubble horizon defined by the distance $H^{-1}$. The significance of the Hubble horizon
is that modes of larger wavelength are frozen while shorter wavelengths oscillate. The value of the co-moving coordinate,
$r_h$,
that characterizes the Hubble horizon is defined by
\be
a r_h = H^{-1}.
\ee
During the initial curvature dominated era, while $H = 1/t$, we find
\be
r_h =1.
\ee
Later, when inflation kicks in, the scale factor and Hubble constant are
 given by
\bea
a \eq t^* e^{H(t-t^*)} \cr
H^2 \eq {\epg V_0}
\eea
and the co-moving coordinate of the Hubble horizon shrinks exponentially,
\be
r_h = e^{H( t^*- t)}.
\ee
Thus, to summarize the history of the Hubble horizon (before reheating), $r_h$ is initially constant
and equal to 1. As potential energy begins to dominate, $r_h $ starts to decrease exponentially. This continues
until reheating, at which point $r_h$ begins to increase. However, prior to reheating the Hubble horizon never exceeds
$r_h =1$. This will play an important role when we discuss the effect of curvature on the large scale structure
of the CMB spectrum.

\setcounter{equation}{0}
\section{Anthropic Bound on Curvature}

Tunnelling of the Coleman-de Luccia sort
produces uniform FRW cosmologies with negative curvature
\cite{Coleman:1980aw}. As we will see in the next section, without observational or anthropic priors, post-tunnelling inflation (if it occurs
at all) would most likely be of  very short duration.  Without a significant period of inflation,
most of the universes residing in $\cal{S}$ will  therefore be dominated by
(negative) curvature and will never have a period of matter or
radiation domination. Thus, before introducing anthropic
observational constraints, the landscape favors an empty, curvature
dominated universe. This, of course, is completely incompatible
with the fact that $\Omega_{total}$ is very close to $1$.

This raises the obvious question of whether there are anthropic
bounds on how large the curvature can be. In other words, does the
formation of life require $\Omega_{total}$ to be close to $1$? We
will show that the  requirement of structure formation (the
anthropic ingredient of the argument) puts an upper bound on the
curvature which is close to the experimental bound. To make the
argument, we will use an analysis that closely parallels
Weinberg's \cite{Weinberg:1987dv}. As in Weinberg's argument, a crucial ingredient will be  to use the
measured value of $\delt$ as an input, and then compute the
maximum values of the cosmological constant and curvature that are
consistent with structure formation. Allowing $\delt$ to vary is a more complicated question which we will not address; see \cite{tegrees}.

\subsection{Calculation}

\label{cal-bound}

A positive cosmological constant is equivalent to a repulsive force that grows linearly with distance.
Negative curvature indicates that the Hubble flow exceeds the Newtonian escape velocity.
It is intuitively obvious that both  positive vacuum energy and negative curvature inhibit structure formation.
If either one is too large, galaxies could not have formed. To be quantitative,
consider a homogeneous,
isotropic, overdense region of the universe of size $ R(t)$ at
the epoch of decoupling. The universe as a whole has some
average matter density $\rho$, curvature $k$, and cosmological
constant $\Lambda$ (we will not include any other form of dark energy).
Locally the magnitude of the overdensity is
$\delta \rho$ and the curvature is $k + \delta k$.
We will ignore the contribution to the density from
radiation, since it is small at decoupling.  The evolution of
$R(t)$ is controlled by the Friedmann equation,
\be
\left( {d R \over dt } \right)^2  + k + \delta k =
{8 \pi G \over 3}  R^2 (\rho + \delta \rho ) + {\Lambda \over 3} R^2 \, .
\lb{friedmann}
\ee
If the background curvature and cosmological constant are zero, the
presence of the overdensity means the region is locally a closed
universe (since $\delta k > 0$) and will collapse.

The matter density satisfies the equation of mass conservation
\be R^3 (\rho + \delta \rho ) = {\cal M},
 \lb{mass-conserv} \ee
where ${\cal M}$ is constant.

The solution of (\ref{friedmann}) is
\be
t =\int {dR \over \sqrt{{8 \pi G \over 3}   {{\cal M} \over R }  + {\Lambda \over 3} R^2
- k - \delta k } }\, .
\lb{sol-fried}
\ee
Expanding the solution for small t gives
\bea
R(t) &\simeq & {\left(6\,\pi G\,{\cal M} \right) }^{1/3}\,t^{2/3}-
\frac{3^{5/3}}{20\, ( 2\, \pi ) ^{1/3}}{k + \delta k \over (G {\cal
M})^{1/3}}\, t ^{4/3} + {\cal O}(t^2) \\
\rho (t) + \delta \rho(t) &\simeq& \frac{1}{6\,G\,\pi \,t^2} + {3^{4/3} \over 40 \,
(2{\cal M})^{2/3}}{k + \delta k \over (\pi G)^{5/3} }\, t^{-4/3}+ {\cal O}(t^{-2/3})
\lb{exp-sol}
\eea
Here $\delta \rho$ is the part of this expression proportional to $\delta k$:
\be
\dro =  {3^{4/3} \over 40 \, (2{\cal M})^{2/3}}{\delta k \over (\pi G)^{5/3} }\, t^{-4/3}  + {\cal O}(t^{-2/3}).
\ee
Assuming $k \sim \delta k$, which is the regime we are interested in, the above expansions are valid so long as $\delta \rho \ll \rho$.  Following Weinberg, we define the parameter $\tilde \rho$ which gives an invariant measure of the strength of the density perturbations:
\be \lb{tro}
\tro \equiv \lim_{t \to 0} {\dro^3 \over \rho^2 } = \left(
{9 \, \delta k \over 40 \, G \, \pi} \right)^3 {1 \over{\cal M}^2 }
\ee
A region with an overdensity $\dro$ will collapse if $( {d R / dt })^2$
becomes zero.  Starting from (\ref{friedmann}), we can minimize with respect to $R$ to obtain
\be
\left( {d R \over dt } \right)^2 \geq  \Lambda^{1/3} (4 \pi G {\cal M})^{2/3} - k - \delta k\, .
\ee
Therefore, the condition for gravitational collapse is
\be
 \Lambda^{1/3} (4 \pi G {\cal M})^{2/3} \leq  k + \delta k\, ,
\ee
which can be rewritten with the use of $\tro $ as
\be
 \Lambda^{1/3} - {k \over  (4 \pi G {\cal M})^{2/3}}  \leq   {10 \over 9}  (4 \pi G \tro)^{1/3}.
\ee
Note that we did not assume that $\delt$ remains small during the evolution to make this analysis.
In a flat universe where $k=0$, the equation above reproduces
Weinberg's condition for structure formation. In an open universe ($k < 0$)
with a \CC $\Lambda >0$, structures will form only if the initial density
perturbations are big enough to overcome the combined effect of the
\CC and the curvature.

We can re-write the bound in terms of the conventional cosmological parameters today:
\be \lb{boundtoday}
{3 \over 2^{2/3}} \, \left( {\Omega_\Lambda \over \Omega_{matter}} \right)^{1/3} + {1 - \Omega_{total} \over \Omega_{matter} }
 \leq   {5 \over 3} \, \left( {\tro \over \rho_0}\right)^{1/3}
\ee
where $\rho_0$ is the matter density today and $\Omega_i$ are evaluated today.
Recall that $\Omega_{matter} =\Omega_{total} - \Omega_{\Lambda}$, so we will think of this as a bound on $\Omega_{total}$ and $\Omega_{\Lambda}$, given a value for $\tro/\rho_0$.

\begin{figure}
\begin{center}
\includegraphics{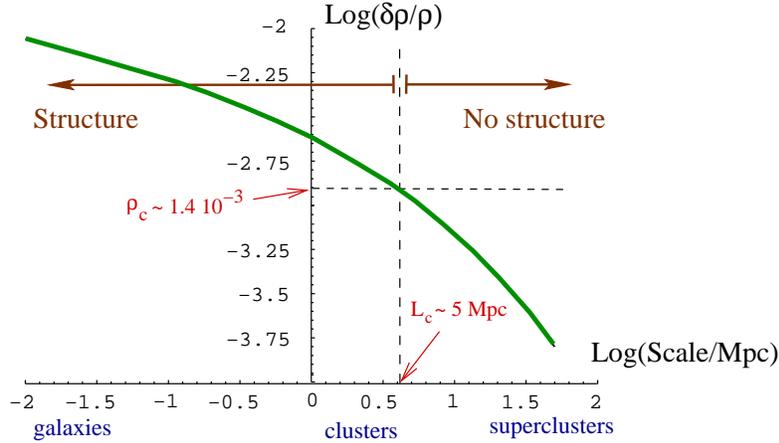}
\caption{Cosmological perturbations at decoupling as a function of comoving scale.  $L_c$
is the critical length below which structure has collapsed or will collapse in our universe, assuming dark energy is a cosmological constant. }
\label{plot-perturb}
\end{center}
\end{figure}

More interesting for our purposes is to express the bound in terms of the radius of curvature at decoupling, $a_{dc}$.
Dropping the cosmological constant term (which only tightens the bound) and setting $\Omega_{matter}=1$ at decoupling,
the bound becomes
\be
(1 - \Omega_{total})_{dc}<  {5 \over 3} \, {\delta \rho \over \rho}
\ee
or
\be
{a_{dc}  H_{dc}} > \left( {5 \over 3} \, {\delta \rho \over \rho}\right)^{-\12}.
\lb{bounddc}
\ee

A complication is that $\delt$ is a function of scale\footnote{We would like to thank Neal Dalal for very helpful discussions on this point.}.
After the generation of perturbations during inflation,
the spectrum evolves due to various effects.  One important
fact, however, is that modes inside the horizon during radiation
domination do not grow significantly.  This means that the power spectrum
at short scales grows only logarithmically.  This is illustrated in Figure \ref{plot-perturb}.

In order to apply our bound (\ref{bounddc}), we should first choose a scale on which to require structure.
If the relevant structures (whose formation we want to guarantee) are typical galaxies, then
the appropriate value of $\delt$ at decoupling is about $10^{-3} $ and (\ref{bounddc}) becomes

\be
{a_{dc} H_{dc}} > 30.
\lb{abound}
\ee

This bound would be slightly relaxed if we only require that dwarf galaxies form. In that case the relevant
value of $\delt$ is about $10^{-2}$ and the anthropic bound loosens by a factor of $3$:
\be
{a_{dc} H_{dc}} > 10.
\lb{dwarf}
\ee
Requiring structure on even shorter scales only changes this slightly, due to the slow growth in $\delt$ mentioned above.

\begin{figure}
\begin{center}
\includegraphics{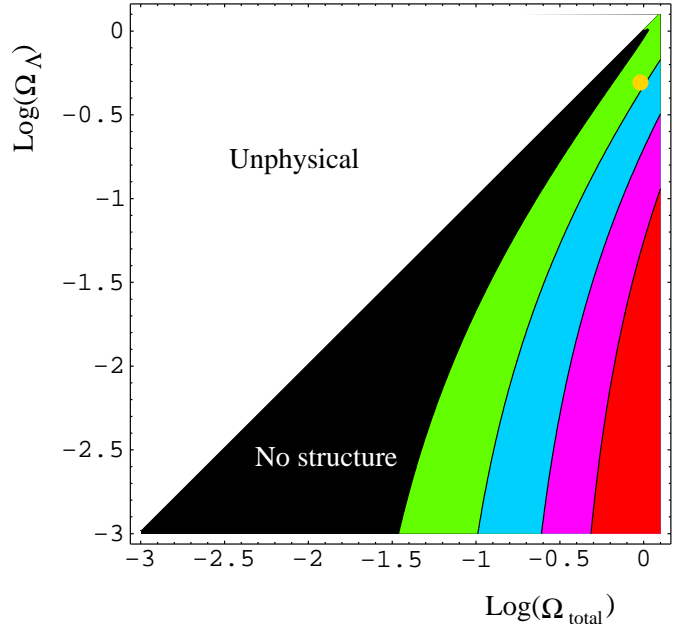}
\caption{Contour lines for the bound, eq. \ref{boundtoday}.  Each point on the diagram corresponds to a different value of $\Omega_{matter} =\Omega_{total} - \Omega_{\Lambda}$; $ (\delta \rho / \rho)_{dc} $ has been set to $3 \times 10^{-3}$, which corresponds to a scale of 1 Mpc. The white triangle corresponds to unphysical universes where $\Omega_{matter} <0$.  In the black area the bound is violated and structure cannot form.  The disk in the upper-right is our universe.}
\label{contourlines}
\end{center}
\end{figure}

Let's compare (\ref{abound}) with the  observational bound. To calculate the observational bound on ${a_{dc} H_{dc}}$
we first use the corresponding bound on $\Omega_{total}$ today.  We will use as a reference number $\Omega_{total} > 0.98$.  This value is about 2 $\sigma$ off the WMAP central value of $1.02 \pm 0.02$ \cite{Bennett:2003bz}, and 1 $\sigma$ off the central value of $0.99 \pm 0.01$ as measured using  large-scale structure surveys \cite{Eisenstein:2005su}.  Using $\Omega_{total} > 0.98$ we obtain

\be
a_0 H_0 > 7.
\lb{todaybound}
\ee
To proceed, recall that $aH = \dot a$ and use the FRW equations of motion
to relate the value of $\dot{a}$ at decoupling to the value today. We find

\be
{\dot{a}_{dc} \over \dot{a}_0}= \sqrt{\Omega_m  (1 + z_{dc}) \left( 1 
+ {1 + z_{dc} \over 1 + z_{eq}} \right) + {\Omega_{\Lambda} \over (1 + z_{dc})^{2}} } \simeq 19,
\lb{dctonow}
\ee
where the $\Omega_i$ are evaluated today and $dc$ and $eq$ stand for decoupling and equality respectively. Combining \ref{dctonow} with \ref{todaybound} we find the
observational bound
\be
{a_{dc} H_{dc}} \simgeq 140.
\lb{obbound}
\ee
The similarity between (\ref{dwarf}) and (\ref{obbound}) is quite striking.

We should also point out that equation \ref{boundtoday}, together with
the scale dependence of $\delt$ (see Figure \ref{plot-perturb}) implies that the bound is {\em not}
satisfied in our universe on scales larger than $L_c \sim 5$ Mpc.
This means that matter on cluster scales and larger today will not continue to collapse.
In fact, due to the effect of the cosmological constant, matter on those scales will actually
dilute and ``inflate away."  Hence we live in the time when the most structure exists.
Similar conclusions were reached in \cite{kleban}.

\setcounter{equation}{0}
\section{Number of Efolds}
\label{Number}
In order to satisfy either the anthropic or observational bounds, the universe
must have undergone a significant period of inflation. Using standard values
for inflation parameters, the minimum number of efolds $N$ needed to satisfy
the observational curvature bound is around $N=62$ \cite{liddleandlyth}. The exact value of this number is logarithmically
sensitive to such parameters as the reheating temperature, the scale of inflation, etc.

The parameter $aH$ depends on $N$ exponentially:
\be
(aH)_{dc} = {(aH)_{dc} \over (aH)_e} \, (aH)_e =  \alpha e^N.
\ee
Here $(aH)_e$ refers to $aH$ at the end of inflation,  $\alpha = (aH)_{dc} / (aH)_e$ is a
factor depending on the expansion of the universe after the end of inflation, and we have defined
$N$ as the number of efolds starting from the first era of curvature domination (when $aH \sim 1$).
We will take $\alpha$ to be of order $e^{-57}$, which corresponds to the value $N =62$ mentioned
above\footnote{For simplicity, we are assuming here that $\alpha$ and $H_{dc}$ do not depend on $N$;
in other words that as we vary the parameters of the inflation potential that determine $N$ we hold fixed
the evolution of the universe after the end of inflation.  Relaxing this assumption should not change the
result significantly.}.
It is quite remarkable that the anthropic
bound (\ref{abound}), thought of as a bound on $N$, is only smaller by about two and a
half efolds than the observational bound \ref{obbound}. This is because $\ln (140/10) \approx 2.5$.
So our bound (obtained by requiring structure on dwarf galaxy scales) is:
\be
N_{structure} > 59.5
\ee
compared with the observational bound (using $\Omega_t > 0.98$)
\be
N_{observed} > 62.
\ee

 We would like to use statistical landscape
arguments \cite{douglas} \cite{douglas2} to estimate the relative number of vacua with such
long-lived inflationary eras and to make some rough statistical guesses
about the actual number of efolds. The duration of inflation is
determined by the parameters appearing in the inflaton potential.
Thus any kind of probabilistic arguments must originate from
statistical assumptions about these parameters.

  The assumption we will use here is that the coefficients in a Taylor
expansion of the potential around generic points are randomly
distributed, over some range of variation that is typically string
scale. We will make the simplest assumption  about the measure, namely it is featureless.
This may be something that can be improved upon by studying the statistics of detailed
string theory models such as \cite{Kachru:2003sx}.

In such a situation, it is of course highly unusual to find
minima in which the \CC is small.  Having found such a minimum, it
is again unlikely that any of the slopes leading down into it are
appropriate for slow roll and long inflation.

We can try to estimate the expected number of efolds $N$
from rolling down such a slope using a simple model. We will assume that the potential is well-approximated
as linear in the
region of interest, so that
\be
V(\phi) = V_0(1 - x {\phi \over \Delta }) , \, \, \, \, \,  \phi \in [\phi_i, \phi_f].
\lb{potential}
\ee
Here $V_0$, $x$, and $\Delta = \phi_f-\phi_i$ are the relevant parameters that we assume vary from $0$ to $1$
with a smooth measure. From now on we will set $8 \pi G = 1$.

The number of efolds is
\be
N = \int H \, dt \simeq { \Delta^2  \over x },
\ee
using the slow roll equations  $3 H \dot \phi \simeq -V'$ and
 $H^2 \simeq   V_0 /3$.
  Finally,
\be
{\delta \rho \over \rho} \sim {H^2 \over \dot \phi} \sim
  {H \Delta \over x} \,  .
\ee
To compute a measure on $N$, we integrate over $V_0$, $x$, and $\Delta$ with the constraint that $\delt $ is fixed:
\bea
P(N) = \int_0^1 dV_0 \, dx \, d\Delta  \, \, \delta ( N - \Delta^2/x ) \, \delta ( (\delt) - V_0^{1/2} \Delta /x) \, F(\Delta, V_0, x) \cr
= {2 (\delt) \over N^4} \int_0^1 d\Delta \, \, \Delta^4 \, F(\Delta, \, (\delt)^2 \Delta^2/N^2, \, \Delta^2/N^2).
\eea
Here $F$ is a measure on the distribution of parameters in the potential.  For a flat distribution ($F=1$) we obtain
\be
P(N) = {2 (\delt) \over 5 N^4}.
\ee

One can use this probability in a number of ways. With no further
constraints, we can estimate the probability that $N$ is greater
than the observational bound $N>62$ (using $\delt \sim 10^{-5}$):
\be
P_{N>62}\sim
\int_{62}^{\infty} P(N) \, dN  \sim  10^{-11.5}.
\ee
This very small probability can be thought of as the flatness problem in this model.

However, if we take into account the anthropic bound ($N>59.5)$ the situation improves dramatically.  In that case
\be
P_{N>62}= {\int_{62}^{\infty} dN/N^3  \over  \int_{59.5}^{\infty} dN/N^3} > 0.88.
\ee

So we see that the probability that the observational bound is satisfied given that structure formed is around 90\%, and we can say that in this model {\em the requirement of structure formation solves the flatness problem.}  Of course the other standard cosmological problems which are solved by sufficient inflation are solved as well.

On the other hand, the probability that $N$ is close to the observational bound is not
insignificant. For example, the probability that $N$ lies in the range between $62$ and $64$
is of order 10\%. This would  correspond roughly to
 $1 - \Omega_{total}$ lying between $0.02$ and $0.0004$, which might be observable in the future.

\setcounter{equation}{0}
\section{ Signatures}
In principle, a tunnelling event in our past would produce observable effects. These effects may or may not
be observable in practice. This depends on whether inflation went on for a long enough duration to
obliterate the effects. The effects we will discuss in this section would  only be observable
if the curvature bound is almost saturated, i.e., if $N$ is close to $62$. The most obvious observable signature of tunnelling
would, of course, be the existence of negative curvature, i.e., $
1-\Omega_{total}> 0$. But it is possible that there are more easily observed effects on the CMB spectrum.

Let us restate the observed bound on curvature ($\Omega_t > 0.98$) in a particularly intuitive way.
What it says is that today, the radius of curvature of
space (the current FRW scale factor $a_0$) is at least 7 times larger than the current Hubble radius $1/H_0= a_0/\dot{a_0}$.
Nevertheless, we will see that there are observable effects on the CMB that can be
very sensitive to the finiteness of the radius of curvature on such scales.

The ratio of the radius of curvature to the Hubble radius is given at all times by $aH$. We can use \ref{obbound} to see
that the ratio at decoupling was even larger: $(aH)_{dc} > 140$. But observations of the CMB today are not constrained by the
size of the horizon at decoupling. As we will see in a moment, the maximum size scales probed by observations of the CMB anisotropy correspond to the
quadrupole (l=2) mode, with a wavelength about 130 times larger than the horizon at decoupling.
Thus, if our universe is close to saturating the observational curvature bound, the lowest l-modes are probing
scales very close to the radius of curvature. In such a case one expects
that the low $l$ multipole moments in the CMB may deviate from scale invariance
in some way.

There are several effects that can distort the flat CMB spectrum at very low $l$. We will mention some
 obvious ones. Let us begin with the metric for open FRW geometries.
 \be
ds^2 = dt^2 -  a(t)^2 \left[dr^2 + (\sinh{r})^2 d\Omega^2 \right].
 \ee

The time $t$ runs from $0$ to $\infty$. The scale factor for small $t$ behaves like
\be
a = t
\ee
during the initial curvature dominated phase. The Hubble parameter, $H$, is $\dot{a
}/a = 1/t$. At this  time the Hubble horizon coordinate, $r_h$, defined by $a r_h = H^{-1} $ is constant in co-moving coordinates:
$r_h = 1$.

At somewhat later time, the vacuum energy $V_0= 3 H_0^2$ will overtake the curvature and the scale factor will
begin to inflate like $a \sim e^{H_0 t}$. This happens when $t \sim H_0^{-1}$. The coordinate, $r_h$, of the Hubble
horizon starts to decrease exponentially and continues to do so until reheating. Thus, before and during
inflation,  $r_h$ is never larger than $r=1$. Perturbations on scales larger than this are frozen and never
equilibrate. They are given by their value on the initial time slice $t=0$.

There are a number of things that suggest that the initial field fluctuations at $r>1$ are very small.
First of all, before the field tunnelled from the previous de Sitter minimum, the inflaton was massive.
Thus there were no long range fluctuations before tunnelling. Secondly, the Coleman De Luccia instanton
in the thin wall approximation leads to a very homogeneous background. Finally, the mode functions on the
negatively curved spatial slices all exponentially tend to zero (with $r$) for $r>1$. All of this
suggests that for large $r$ the initial fluctuations are exponentially small. The $l=2$ mode
is probing comoving coordinate scales that are at most order $1$ (see eq. \ref{lscale}), and the higher $l$ modes probe even smaller $r$.
Thus one might expect that the trend is for the low $l$ multipoles to be small.

The scale $r=1$ is the comoving radius of curvature of the spatial slices.  Let us again assume that $\Omega_{total} = 0.98$, and ask what scale the $l=2$ mode is associated with.  This scale is the diameter of the last-scattering sphere.  Integrating back along the photon geodesic gives
\be\label{lscale}
r_{l=2} = 2 \int_{t_{dc}}^{t_0} {dt \over a} = 
2 \int_{1 / (1 + z_{dc})}^1 \, \, {dx \over x \sqrt{1 + x^2 (a_0 H_0)^2 \left( \Omega_m x^{-3} + 
\Omega_{\Lambda} + \Omega_r x^{-4}\right) }} = 0.95.
\ee
So we see that the wavelength which makes the largest contribution to the $l=2$ mode of the CMB is almost exactly the same as the radius of curvature.

However there is a possibility that the suppression effects may extend to distances somewhat smaller than the
radius of curvature. Inflation only begins when curvature dominance gives way to vacuum dominance. At that time there is no
reason to expect significant fluctuation. It takes time for de Sitter space to come to thermal equilibrium.
A rough estimate is that the time scale for equilibration is of order
\be
t_{eq}\sim H^{-1}\log{m_p / H},
\ee
where $m_p$ is the Planck mass \cite{Kaloper:2002cs}. This can correspond to a significant number of efolds, since $H < 10^{-5} m_p$.  Any modes that leave the horizon before this time may still be suppressed.

Another interesting effect involves the shape of the potential at the onset of inflation. We have assumed a linear
potential after the tunnelling event but that is obviously not correct for times shortly after tunnelling.
What we should expect is a steeper curved potential that descends from the tunnelling point to the flat inflationary
plateau. Let us model this by introducing a quadratic term in the potential for negative $\phi$.\footnote{We have shifted conventions so that $\f(0) < 0$.} Thus for $\phi > 0$
we have the usual linear potential in  \ref{potential},  but for $\phi<0$ the potential has the form
\be
V= V_0 -x V_0 { \phi \over \Delta \phi} + {m^2 \phi^2 \over 2  }.
\ee

The low $l$ modes in the CMB receive their largest contributions from long wavelength perturbations.  Let us assume that at the time when these perturbations were leaving the horizon that
the field was negative but close to zero. Then the fluctuations generated at this time
are given by
\be
{\delta \rho \over \rho}= {V^{3/2} \over V'  } = {V_0^{1/2} \Delta \phi \over x}
(1 - ({3x \over 2\Delta \phi }- {m^2 \Delta \phi\over x V_0} )\phi  ) +O(\phi^2).
\ee

Evidently the sign of the effect depends on whether $m$ is bigger or smaller than ${x \over \Delta \phi}  V_0^{1/2}$. In particular
 for $m > {x \over \Delta \phi}  V_0^{1/2}$ the power in these modes would be suppressed. For the choice
 \bea
V_0 &=& 10^{-13}\cr
x&=& 10^{-2}
 \eea
 we find that for $m > 10^{-8.5}m_p$, the primordial power spectrum at long wavelengths would be suppressed.  However, this may or may not correspond to a suppression in total power in the low $l$ CMB, since effects such as the late integrated Sachs-Wolfe effect can in some models overcompensate for the suppression \cite{lowlanomaly}.

\section{Conclusions}
This paper is based on one principle and two surprising numerical facts. The principle is that our
universe was created by a tunnelling event from a neighboring valley of the landscape. From this it follows
that we live in an open FRW universe with special smooth initial conditions.

The numerical facts that we found surprising, and seem not to be well known, are the following:

\begin{itemize}

\item Anthropic considerations similar to Weinberg's constrain the curvature to be small. In terms of efolds,
the existence of galaxies requires $N$ to be at least within two efolds of the observational bound.

\item If saturated, the observational bounds on curvature are equivalent to the statement that the 
$l=2$ mode of the CMB anisotropy corresponds to a scale equal to the radius of curvature of the universe.

\end{itemize}

We have also given an admittedly crude argument that landscape statistics may favor a small number of
efolds. Obviously there is a great deal of work that can be done to refine this idea.
Models of the kind studied by \cite{Kachru:2003sx} could potentially provide statistical distributions of parameters.

In view of these points, our opinion is that  the observed
suppression of the quadrupole and octopole anisotropies, seen by
WMAP, should not be quickly dismissed as cosmic variance accidents. However (as several authors have pointed
out \cite{lowlanomaly}) a long wavelength suppression of the primordial spectrum can not explain the entire suppression of the observed quadupole/octopole
because of the enhancement due to the late time integrated Sachs Wolfe effect (which would be
 unchanged in the models considered in this paper).

\section{Acknowledgements}\nonumber
All four of us would like to thank the faculty and staff of the Korea Institute for
Advanced Study, where this work was
initiated, for their hospitality.  MK and MRM would like to thank the University of Texas, Austin for hospitality during the completion of the work.  We thank Gia Dvali, Willy Fischler, Gregory Gabadadze, Kimyeong Lee, Andrei Linde, Brice Menard, Sonia Paban, Carlos Pe\~na-Garay, Ra\'ul Rabad\'an, Steve Shenker, Scott Thomas, Larus Thorlacius, Bob Wagoner, and Neal Weiner for useful discussions.   LS is particularly grateful to Willy Fischler and Sonia Paban for important discussions.  MK would like to especially thank Neal Dalal for very useful conversations and for providing a copy of the program sigma.

The work of M.K. is supported by NSF grant PHY-0070928, and that
of L.S. by NSF grant PHY-0097915.

\end{document}